%% file: main_for_arxiv.tex
\documentclass[aps, pra, superscriptaddress, twocolumn, amsfonts, amsmath, amssymb, floatfix]{revtex4-2}
\usepackage{graphicx}
\usepackage{float}
\usepackage{sidecap}
\usepackage{dcolumn}
\usepackage{bm}
\usepackage{epsfig}
\usepackage{braket}
\usepackage{color}
\usepackage{ulem}
\usepackage{amsmath}
\usepackage{wasysym}
\usepackage[colorlinks=true, letterpaper=true, pdfstartview=FitV, linkcolor=blue, citecolor=blue, urlcolor=blue]{hyperref}
\usepackage{orcidlink}
\usepackage{lipsum}

\begin{document}

\title{Bose-Einstein condensates in a spin-twisted harmonic trap}

\author{Huaxin He~\orcidlink{0000-0003-4067-191X}}
\thanks{These authors contributed equally to this work.}
\affiliation{Provincial Key Laboratory of Multimodal Perceiving and Intelligent Systems, Jiaxing University, Jiaxing, 314001, China}
\affiliation{Institute for Quantum Science and Technology, Department of Physics, Shanghai University, Shanghai 200444, China}

\author{Fengtao Pang~\orcidlink{0000-0002-2784-0064}}
\thanks{These authors contributed equally to this work.}
\affiliation{Institute for Quantum Science and Technology, Department of Physics, Shanghai University, Shanghai 200444, China}

\author{Xianchao Zhang}
\affiliation{Provincial Key Laboratory of Multimodal Perceiving and Intelligent Systems, Jiaxing University, Jiaxing, 314001, China}

\author{Yongping Zhang~\orcidlink{0000-0001-5090-9173}}
\email{yongping11@t.shu.edu.cn}
\affiliation{Institute for Quantum Science and Technology, Department of Physics, Shanghai University, Shanghai 200444, China}

\author{Chunlei Qu~\orcidlink{0000-0002-3080-8698}}
\email{cqu5@stevens.edu}
\affiliation{Department of Physics, Stevens Institute of Technology, Hoboken, New Jersey 07030, USA}
\affiliation{Center for Quantum Science and Engineering, Stevens Institute of Technology, Hoboken, New Jersey 07030, USA}

\begin{abstract}
We investigate the ground-state phases and spin-scissors dynamics of binary Bose–Einstein condensates confined in a twisted two-dimensional harmonic trap. The ground state hosts three distinct phases—phase-separated, polarized, and phase-mixed—determined by the Rabi coupling, interaction ratio $G$ (between inter-component and intra-component interactions), and spin-twisting which induces edge-localized polarization through position-dependent detuning. In the phase-mixed regime, the ground state is characterized by a finite spin-scissors susceptibility and can be accurately described using local density approximation. In the dynamics, the system exhibits stable periodic beating in the phase-mixed state for $G \leq 1$. For $G > 1$, its evolution progresses from beat damping (phase-separated state) to polarized relaxation (polarized state), finally reaching stable periodic beating (phase-mixed state) after a finite waiting time. The dependence of the waiting time contrasts sharply with the monotonic behavior of one-dimensional spin-dipole dynamics, revealing qualitatively distinct mechanisms governed by geometry and interactions. In summary, these results establish a unified link between ground-state properties and nonequilibrium responses in twisted spinor condensates, offering a versatile platform for exploring spin-related quantum many-body phenomena.
\end{abstract}

\maketitle

\section{Introduction}

In recent years, significant advances have been made in twistronics, particularly in the study of twisted bilayer graphene systems that exhibit fascinating strongly correlated phenomena~\cite{bistritzer2011moire,cao2018correlated,cao2018unconventional,andrei2020graphene,torma2022superconductivity}. By modifying electronic transport~\cite{PhysRevLett.121.257001,PhysRevX.8.041041,PhysRevLett.122.257002,choi2019electronic,yankowitz2019tuning}, renormalizing the Fermi velocity~\cite{PhysRevB.92.201408}, and inducing flat bands at magic angles~\cite{codecido2019correlated,balents2020superconductivity,PhysRevLett.122.106405}, these phenomena create a unique platform for discovering novel quantum states~\cite{sharpe2019emergent,chen2019evidence,PhysRevLett.123.197702,lu2019superconductors,cao2020tunable,nuckolls2020strongly}. The twist-engineering concept has further been extended to artificial quantum systems beyond electronic materials. This is exemplified in photonic crystals, where introducing a twist (or Moiré patterns) enables photonic localization and flat-band physics~\cite{wang2020localization,PhysRevLett.126.223601}. Moreover, twist engineering has also been leveraged to explore novel effects in low dimensions, such as realizing two-dimensional (2D) topological pumping and probing nonlinear phenomena in twisted settings~\cite{wang2022two,PhysRevLett.127.163902,du2024nonlinear}.

Owing to their exceptional quantum coherence and precise tunability, ultracold atomic Bose-Einstein condensates (BECs) serve as an invaluable platform for quantum simulation of condensed matter systems, enabling exploration of complex quantum phenomena under exquisitely controlled conditions~\cite{pitaevskii2003bose,RevModPhys.80.885,lewenstein2012ultracold,schafer2020tools}. Recently, theoretical approaches utilizing spin-dependent optical lattices and synthetic dimension have been proposed for realizing tunable twisted bilayer configurations~\cite{grass2016proximity,PhysRevA.100.053604,PhysRevLett.125.030504,PhysRevLett.126.103201,lee2022emulating,PhysRevA.111.023320,zpqd-ryjm,liang2025atomic}. Experimentally, precise control of interlayer tunneling in twisted square-lattice bilayers via microwave fields has facilitated the observation of Moiré patterns in BECs and revealed novel quantum phases across the superfluid–Mott insulator transition~\cite{meng2023atomic}. This ultracold-atom platform exhibits unique capabilities: it permits dynamic tuning of both twist angles and interlayer coupling strengths while achieving the first quantum simulation of 3D twisted geometries~\cite{PhysRevLett.133.163401}. Further studies demonstrate that the geometric Moiré effect induces Hofstadter butterfly-like fractal spectra~\cite{wan2024fractal}, accompanied by interaction-driven emergent phenomena including rich dynamics and topological phases~\cite{PhysRevA.107.033316,PhysRevResearch.6.L042066,PhysRevB.111.024511,paul2025interacting}. Similar to the twisted lattice systems mentioned above, spin-dependent harmonic trap also introduces novel physics to BECs. In 1D scenarios, the introduction of spin-dependent harmonic trap enables the exploration of collective excitations related to spin-dipole modes, offering valuable opportunities for investigating spin dynamics~\cite{sartori2015spin}. When interactions are considered, phenomena such as phase separation and phase mixing significantly influence the ground-state structure and spin-dipole oscillations~\cite{PhysRevA.94.063652,sartori2015spin}. Extending this concept to 2D systems, one can generate a spin-twisted harmonic trap, which provides a platform with new degrees of freedom to explore exotic spin dynamics, such as spin-scissors modes. However, a comprehensive understanding of the ground states and dynamics in such a spin-twisted trap has yet to be established.

This work investigates the ground-state phases and spin-scissors dynamics of two-component Rabi coupled BECs confined in a twisted 2D harmonic trap. We first map out the phase diagrams for varying Rabi coupling and interaction ratio $G$ (between inter-component and intra-component interactions), both with and without spin-twisting, and classify the corresponding ground states into three distinct phases: phase-separated, polarized, and phase-mixed. We find that integrated and local spin polarization can be used to distinguish these phases, while spin-twisting induces edge-localized polarization in the condensate. For the phase-mixed regime, an approximate analytical expression for the spin-scissors susceptibility is derived, showing excellent agreement with numerical simulations. Dynamical simulations reveal qualitatively different behaviors across the three phases, including stable periodic beating for $G \le 1$ and the emergence of spontaneous polarization with a finite waiting time for $G>1$. The results are contrasted with 1D spin-dipole dynamics, highlighting the distinct mechanisms introduced by 2D geometry and spin-twisting.

The paper is organized as follows. Section~\ref{sec2} presents the model of the BECs in a spin-twisted harmonic trap. In Sec.~\ref{sec3}, the ground-state properties of the system are analyzed using numerical methods and analytical derivations. In Sec.~\ref{sec4}, we discuss the spin-scissors dynamical behaviors of the system in different parameter regimes. Finally, in Sec.~\ref{sec5}, we summarize our conclusions.

\section{Binary BECs Confined in a Spin-Twisted harmonic trap}\label{sec2}

\begin{figure}[t!]
\centerline{
\includegraphics[width=0.20\textwidth]{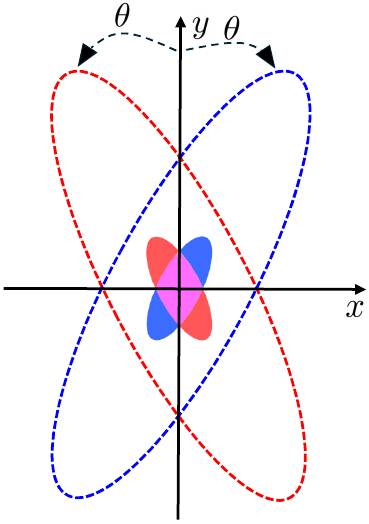}}
\caption{Schematic illustration of a BEC in the spin-twisted harmonic trap. The dashed lines indicate the shape of the external potential after being twisted by an angle $\theta$, clockwise for the spin-up $\mid\uparrow\rangle$ component and counterclockwise for the spin-down $\mid\downarrow\rangle$ component. The red and blue regions represent the atomic clouds of $\mid\uparrow\rangle$ and $\mid\downarrow\rangle$, respectively. The spin-twisting angle is exaggerated for visual clarity.
}
\label{fig1}
\end{figure}

We investigate a binary mixture of BECs confined in a spin-twisted harmonic trap at zero-temperature. The two components are coupled by the Rabi coupling and interact via two-body collisions~\cite{abad2013study,PhysRevA.91.063635,PhysRevA.95.033614,PhysRevA.96.063623,PhysRevA.103.053322,recati2022coherently}. Within the mean-field framework, the system dynamics are governed by the Gross-Pitaevskii equation (GPE)
\begin{equation}
\label{GPE1}
i\hbar\frac{\partial\psi_j}{\partial t}=\left[-\frac{\hbar^2\nabla^2}{2m}+V_j+\sum_{k = 1,2}g_{jk}|\psi_k|^2\right]\psi_j-\Omega\psi_{3 - j},
\end{equation}
where $\psi_j$ ($j = 1, 2$) are the order parameter of component-$j$ and $m$ is the atomic mass. The nonlinear interaction strength $g_{jk}=4\pi\hbar^2a_{jk}/m$, with $a_{jk}$ being the $s$-wave scattering length between component-$j$ and -$k$, which can be adjusted over a wide range by Feshbach resonance~\cite{RevModPhys.82.1225}. Here we assume that the interaction is repulsive such that all interaction strengths $g_{jk} > 0$. Moreover, for simplicity, we set $g_{11} = g_{22} = g$ for intra-component interactions, while for inter-component interactions it satisfies $g_{12} = g_{21}$. To study the interplay between intra- and inter-component interactions, we define an interaction ratio, namely $G = g_{12}/g$. The parameter $\Omega$ denotes the Rabi coupling strength, which is experimentally implemented via radio-frequency or microwave fields~\cite{PhysRevLett.100.150401,PhysRevLett.111.264101}. The Rabi coupling controls population balance by driving coherent mixing between the two components~\cite{abad2013study}, which locks their relative phase and breaks the conservation of the relative particle number ($N_1-N_2$). As a result, it eliminates one of the two Goldstone modes from the excitation spectrum. Only the total particle number, $N = N_1 + N_2$, remains conserved, and the two components share the same chemical potential, i.e., $\mu_1 = \mu_2 = \mu$~\cite{PhysRevA.55.2935,PhysRevA.67.023606}. The harmonic trap potential for component $j$ in the $x$-$y$ plane is given by $V_j = \dfrac{1}{2}m\omega_x^2 x_j^2 + \dfrac{1}{2}m\omega_y^2 y_j^2$. For a spin-twisting angle of $\theta = 0^\circ$, the two components share the same coordinate system ($x_1 = x_2$, $y_1 = y_2$). To introduce a spin-twist, we rotate the harmonic traps for the two components by angles of $+\theta$ and $-\theta$, respectively, about the $y$-axis, as illustrated in Fig.~\ref{fig1}. This rotation transforms the coordinates for each component as follows:
\begin{equation}
    \begin{aligned}
    x_1& = x\cos\theta - y\sin\theta, &y_1& = y\cos\theta + x\sin\theta,\\
    x_2& = x\cos\theta + y\sin\theta, &y_2& = y\cos\theta - x\sin\theta.
    \end{aligned}
\end{equation}
Through this transformation, the system's external potential can be rewritten as  
\begin{equation}
\label{V1}
    V = \frac{1}{2}m\bar{\omega}_x^2 x^2 + \frac{1}{2}m\bar{\omega}_y^2 y^2 - \eta xy\sigma_z,
\end{equation}  
where the spin-twisting renormalizes the harmonic trap, producing effective trapping frequencies along the $x$- and $y$-directions:
$\bar{\omega}_x = \sqrt{\omega_x^2 \cos^2\theta + \omega_y^2 \sin^2\theta}$ and
$\bar{\omega}_y = \sqrt{\omega_x^2 \sin^2\theta + \omega_y^2 \cos^2\theta}$. More significantly, the spin-twisting additionally generates a position-dependent detuning term proportional to $xy$, with detuning strength $\eta = m\sin(2\theta)(\omega_x^2 - \omega_y^2)/2$. This strength depends on both the rotation angle and the anisotropy of the harmonic trap—a mechanism fundamentally distinct from the twist implementations in optical lattices.

To investigate the ground-state properties and dynamical behaviors of BECs in a spin-twisted harmonic trap, we numerically solve the GPE and employ the local density approximation (LDA)—which involves neglecting the kinetic (quantum pressure) term of GPE—for analytical insights~\cite{sartori2015spin}. For computational convenience, we rewrite Eq.~\eqref{GPE1} in dimensionless form. In this work, we use $k_0 = 7.95 \times 10^6\ \text{m}^{-1}$ as the momentum unit, which correspondingly defines the units of length, energy, and time as $k_0^{-1}$, $E_0 = \hbar^2 k_0^2/(2m)$, and $\hbar/E_0$, respectively. Our numerical simulations are conducted on a $256 \times 256$ grid in both the $x$ and $y$ directions, with a time step of $\Delta t = 0.01$ to ensure numerical stability. The confinement frequencies of the harmonic trap are set to $(\omega_x, \omega_y) = 2 \pi \times (175, 75)$ Hz. For the consistency between the numerical results of the GPE and the analytical results of the LDA, the total number of atoms is fixed at $N = 5.0 \times 10^{4}$.

\section{Density Distributions and Spin-Scissors Susceptibility of Ground States}
\label{sec3}
\subsection{Phase Diagram}

\begin{figure}[t]
\centerline{
\includegraphics[width=0.5\textwidth]{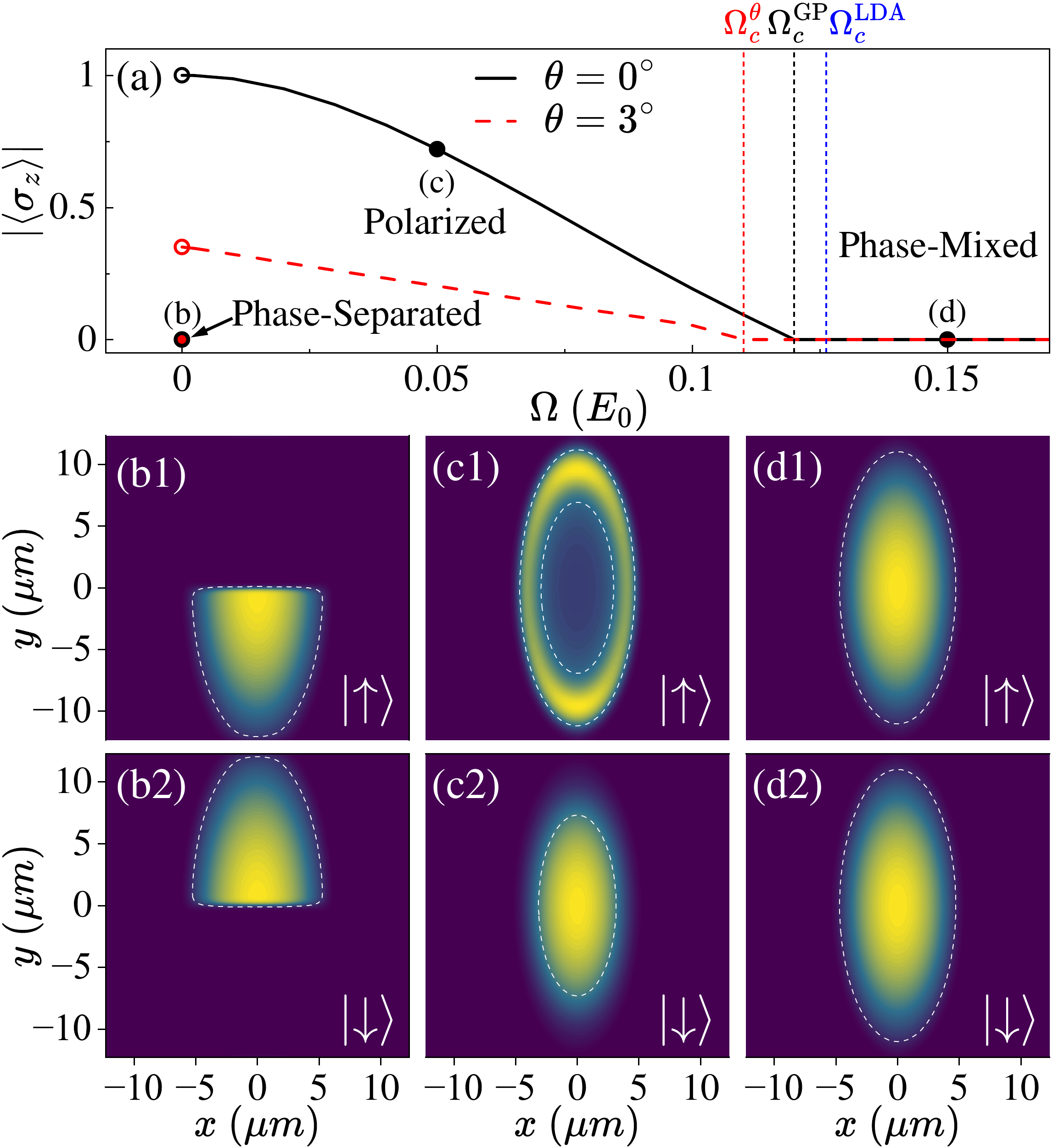}}
\caption{(a) The absolute value of integrated spin polarization $|\langle \sigma_z \rangle|$ as a function of the Rabi coupling strength $\Omega$ for interaction ratio $G = 1.3$. The black solid curve shows the GPE simulation results for a spin-twisting angle $\theta = 0^\circ$, while the red dashed line represents the corresponding results for $\theta = 3^\circ$. Hollow circles at the left endpoints indicate the absence of data. Both cases yield $|\langle \sigma_z \rangle|=0$ at $\Omega=0E_0$, marked by black and red spheres for $\theta=0^\circ$ and $\theta=3^\circ$ respectively. The blue vertical dashed line indicates the critical Rabi coupling $\Omega_c^\text{LDA} \approx 0.126E_0$ predicted by the LDA, marking the transition from polarized states to phase-mixed states. The black vertical dashed line shows the critical Rabi coupling $\Omega_c^\text{GP}$ obtained from GPE simulations for the untwisted system, while the red vertical dashed line denotes the corresponding critical value $\Omega_c^\theta$ for the twisted system. (b–d) Density profiles of the two components in three distinct phases for $\theta = 0^\circ$: phase-separated, polarized, and phase-mixed. The white dashed lines in each panel indicate the effective boundary of the BECs.
}
\label{fig2}
\end{figure}

\begin{figure*}[t]
\centerline{
\includegraphics[width=1\textwidth]{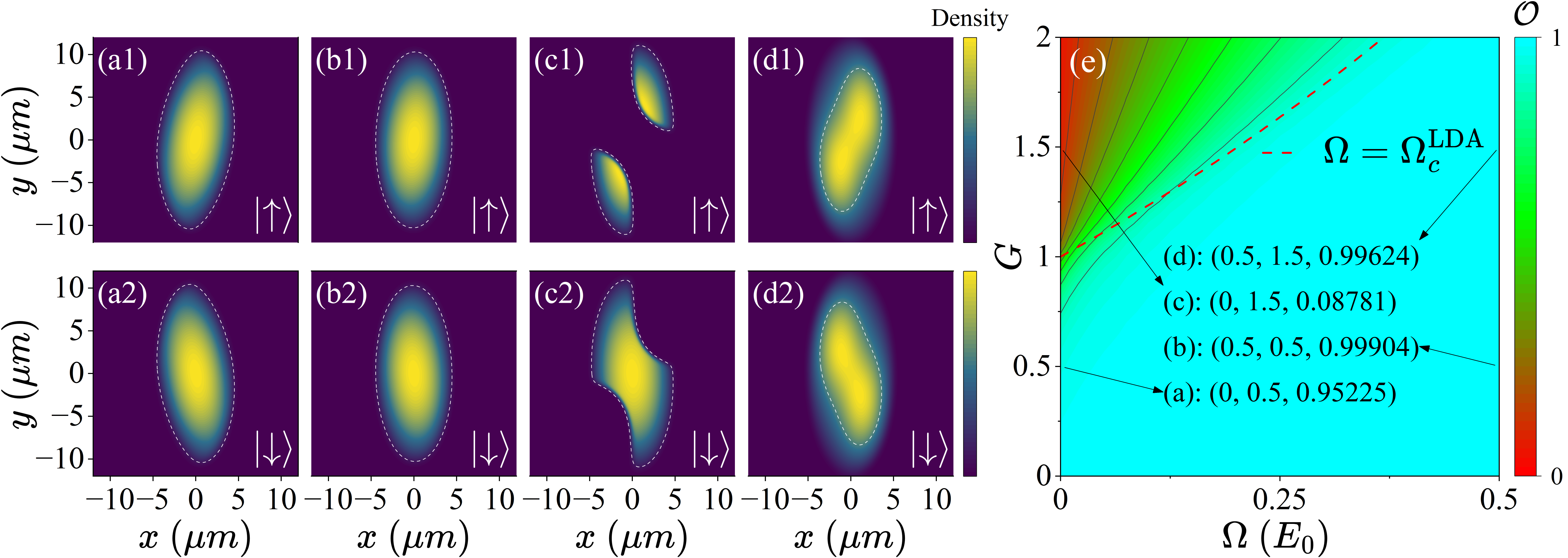}}
\caption{(a–d) Ground-state density distributions of binary BECs under varying interaction ratios $G$ and Rabi coupling strengths $\Omega$ with a spin-twisting angle $\theta = 3^\circ$: (a1, a2) $G = 0.5$ and $\Omega = 0E_0$; (b1, b2) $G = 0.5$ and $\Omega = 0.5E_0$; (c1, c2) $G = 1.5$ and $\Omega = 0E_0$; and (d1, d2) $G = 1.5$ and $\Omega = 0.5E_0$. The white dashed lines in each panel indicate the effective boundary of the BECs. (e) Contour map of the overlap order parameter $\mathcal{O}$ as a function of $\Omega$ and $G$, with the spin-twisting angle $\theta = 3^\circ$. The points labeled (a)–(d) correspond to the parameter sets used in (a–d), and their coordinates and the overlap $(\Omega, G, \mathcal{O})$ are annotated accordingly.
}
\label{fig3}
\end{figure*}

The ground state of the system can be obtained through the imaginary time evolution of Eq.~\eqref{GPE1}. We first examine the case with zero spin-twisting angle ($\theta = 0^\circ$), where the competition between inter- and intra-component interactions, quantified by $G$, together with the Rabi coupling $\Omega$, determines the type of ground state configuration. For $G \le 1$ and $\Omega \geq 0E_0$, the ground state is a fully miscible state—referred to hereafter as the phase-mixed state—characterized by both zero integrated spin polarization ($\langle \sigma_z \rangle \equiv (N_\uparrow - N_\downarrow)/N = 0$) and zero local spin polarization ($s_z = n_1 - n_2 = 0$). When $G > 1$, the interaction can be rewritten as $\frac{1}{2}g(1+G)(|\psi_1| ^2 + |\psi_2| ^2) I + \frac{1}{2}g(1-G)( |\psi_1| ^2 - |\psi_2| ^2)\sigma_z$, which shows that the negative coefficient energetically stabilizes a polarized state~\cite{PhysRevA.76.053615,abad2013study}. Under harmonic confinement, the system adopts a Thomas–Fermi (TF) density profile that decays radially from the trap center. As a result, interaction-induced polarization effects are most pronounced in the central region of the condensate. As shown by the black solid curve for $\theta = 0^\circ$ in Fig.~\ref{fig2}\hyperref[fig2]{(a)}, the behavior of the total spin polarization $|\langle \sigma_z \rangle|$ can be divided into three regions corresponding to three distinct ground states. First, in the absence of Rabi coupling ($\Omega = 0E_0$) and for $G > 1$, the system forms a phase-separated state, which exhibits zero integrated spin polarization ($\langle \sigma_z \rangle = 0$) but nonzero local spin polarization ($s_z \ne 0$). This structure, visible in Fig.~\ref{fig2}\hyperref[fig2]{(b1,b2)}, features a domain wall separating the two components. Introducing a finite Rabi coupling ($\Omega > 0E_0$) drives the system into a polarized phase with both nonzero integrated ($\langle \sigma_z \rangle \ne 0$) and local ($s_z \ne 0$) spin polarization. This is illustrated in Fig.~\ref{fig2}\hyperref[fig2]{(c1,c2)}, where the densities of the two components differ significantly in the center. As $\Omega$ increases further, the integrated spin polarization decreases and mixing is enhanced, eventually leading to a phase-mixed state with zero integrated spin polarization, as shown in Fig.~\ref{fig2}\hyperref[fig2]{(d1,d2)}. We define $\Omega_c$ as the critical Rabi coupling strength at which the system transitions from the polarized to the phase-mixed state. The vertical black dashed line in Fig.~\ref{fig2}\hyperref[fig2]{(a)} indicates the critical value $\Omega_c^\text{GP}$ obtained from GPE simulations for the system without spin-twisting~\cite{sartori2015spin}. Using the LDA, an analytical expression for $\Omega_c$ is derived, marked by the vertical blue dashed line labeled $\Omega_c^\text{LDA}$.

For non-zero spin-twisting angles ($\theta \neq 0^\circ$), the position-dependent detuning in Eq.~\eqref{V1} becomes significant. Its magnitude is largest at the periphery of condensates, leading to edge-dominated phase separation. This mechanism contrasts with the center-driven phase separation induced by inter-component interactions when $G > 1$. For $G<1$ and $\Omega \geq 0$, the condensates remain in the phase-mixed state. However, unlike the untwisted case, the ground state develops finite local spin polarization ($s_z \neq 0$) near the edges while maintaining a vanishing integrated spin polarization ($\langle\sigma_z\rangle = 0$), see Figs.~\ref{fig3}\hyperref[fig3]{(a,b)}. As $G$ increases toward $1$, this edge effect becomes more pronounced; increasing $\Omega$, on the other hand, enhances miscibility and suppresses the edge effect. For $G>1$, Figure~\ref{fig2}\hyperref[fig2]{(a)} also shows that the three characteristic phases—phase-separated, polarized, and phase-mixed—still persist at $\theta \neq 0^\circ$. Nevertheless, compared with the untwisted case, spin twisting reduces the extent of the polarized domain (red dashed curve), shifts the transition point $\Omega_c^\theta$ to smaller values, and thereby lowers the integrated spin polarization. In the phase-separated state, the edge effect induced by spin twisting is most representative: the local spin polarization manifests as one component occupying two spatially separated regions, which are divided by a domain wall from the other component, as shown in Fig.~\ref{fig3}\hyperref[fig3]{(c)}. For the polarized state, spin twisting occurs between the two components on the basis of partial miscibility. In the phase-mixed state, the Rabi coupling ($\Omega > \Omega_c^\theta$) mixes the two components so that their overall spatial profiles nearly coincide. However, due to the strong inter-component interaction ($G>1$), the effect of spin-twisting remains manifest inside the density distributions, where clear distortions are still visible [Fig.~\ref{fig3}\hyperref[fig3]{(d)} vs. Fig.~\ref{fig3}\hyperref[fig3]{(b)}].

The emergence of local spin polarization at the BECs edges motivates us to introduce an overlap order parameter,
\begin{equation}
\label{Over1}
\mathcal{O} = \frac{1}{\sqrt{N_1 N_2}}\iint \sqrt{n_1 n_2} \, d\mathbf{r},
\end{equation}
which quantifies the degree of miscibility between the two components. In Eq.~\eqref{Over1}, $ n_i \equiv |\psi_i|^2 $ is the density distribution and $ N_i $ is the particle number of each component. In this description, for the zero spin-twisting angle ($\theta = 0^\circ$) the phase-separated state corresponds to $\mathcal{O} \approx 0$, the phase-mixed state to $\mathcal{O} \approx 1$, and the polarized state exhibits intermediate values $\mathcal{O} \in (0,1)$. However, for a non-zero spin-twisting angle ($\theta \ne 0^\circ$), the presence of local spin polarization at the edges causes the overlap value in the phase-mixed state to deviate significantly from unity. As shown in Fig.~\ref{fig3}\hyperref[fig3]{(e)}, this deviation manifests in two regions. First, in the regime where $G<1$ but close to $G=1$, and for relatively small $\Omega$, a chroma-transition zone still appears, since the edge-dominated phase separation can be not fully compensated by $\Omega$. Second, for $G>1$ near $\Omega \simeq \Omega_c$, the system undergoes a transition from the polarized state to the phase-mixed state. To characterize this deviation from unity caused by local spin polarization, we plot the critical value $\Omega_c^\text{LDA}$ predicted by the LDA (red dashed curve), and the crossing between the chroma-transition contour and this reference line is clearly visible.

This establishes a complete spatial landscape: $G>1$ favors center-driven phase separation, $\Omega$ promotes global uniformity across the system, and spin-twisting induces edge-dominated phase separation. In addition, spin-twisting gives rise to local spin polarization at the edges, which is the key feature for understanding its effect.

\subsection{Local Density Approximation}
To develop an analytical description of the ground state, the condensate wave function is parameterized as $\psi_j = \sqrt{n_j} \exp(i\phi_j - i\mu t)$, where $n_j(x,y)$ and $\phi_j(x,y)$ denote the local density and phase of component-$j$, respectively. Substituting the ansatz into Eq.~\eqref{GPE1} and invoking LDA yields a simplified form of the GPE. In this work, the Rabi coupling is introduced as $-\Omega \sigma_x$, which locks the relative phase to $\phi_1 - \phi_2 = 0$. Under this constraint, Eq.~\eqref{GPE1} reduces to a set of coupled equations~\cite{sartori2015spin,abad2013study}
\begin{subequations}
\begin{align}
        \left(g - g_{12}+\frac{\Omega}{\sqrt{n_1n_2}}\right)(n_1 - n_2)&=V_2 - V_1,\label{eq5a}\\
        \left(g + g_{12}-\frac{\Omega}{\sqrt{n_1n_2}}\right)(n_1 + n_2)&=2\mu - V_2 - V_1.\label{eq5b}
\end{align}
\end{subequations}
These equations admit analytical solutions for the ground-state density profiles in specific cases. In particular, when the spin-twisting angle $\theta$ is sufficiently small for the TF approximation to hold, we can obtain analytical solutions.

For the specific case of the zero spin-twisting angle $\theta = 0^\circ$, the external potentials are equal ($V_1 = V_2$). Under this condition, when the system is in the phase-mixed state, the densities satisfy $n_1 = n_2$ throughout the entire condensate. Under these conditions, Eqs.~\eqref{eq5a} and~\eqref{eq5b} yield a symmetric TF profile for each component~\cite{pitaevskii2003bose}:
\begin{equation}
\label{HD4}
    n_{1,2}\equiv n_{1,2}(x,y)=\frac{\mu+\Omega}{g + g_{12}}\left(1-\frac{x^2}{R_{x}^2}-\frac{y^2}{R_{y}^2}\right),
\end{equation}
where the TF radii are given by $R_\nu = \sqrt{2(\mu + \Omega)/m\omega_\nu^2}$ for $\nu = x,y$. The chemical potential $\mu$ is determined by enforcing the normalization condition $N = \iint (n_1 + n_2) \, d\mathbf{r}$, yielding
\begin{equation}
\label{HD5}
    \begin{aligned}
        \mu&=\left[\frac{m N \omega_x\omega_y(g_{12}+g)}{2\pi}\right]^\frac{1}{2}-\Omega.
    \end{aligned}
\end{equation}

Furthermore, for $G>1$, as the Rabi coupling increases, the ground state transitions from the phase-separated state to the polarized state and eventually, beyond a critical value $\Omega_c$, to the phase-mixed state. This evolution is illustrated in Fig.~\ref{fig2}\hyperref[fig2]{(b)} to Fig.~\ref{fig2}\hyperref[fig2]{(d)}. In the polarized state, a polarized core always exists where $n_1(X,Y) \ne n_2(X,Y)$, with $(X,Y)$ denoting positions within this core region. There, the inter-component interaction satisfies the condition $g_{12} > g + 2\Omega / n(X,Y)$, where $n(X,Y) = n_1 + n_2$ is the total density, as derived from Eq.~\eqref{eq5a}. Note that if this condition is not met at the trap center, the entire system remains unpolarized—i.e., in the phase-mixed state. This allows us to define a critical Rabi coupling strength~\cite{sartori2015spin}:
\begin{equation}
    \Omega_c = \frac{1}{2}n(0,0)(g_{12} - g),
\end{equation}
where $n(0,0)$ denotes the total density at the center of the trap. This expression always remains nonzero for $G > 1$. As shown in Fig.~\ref{fig2}\hyperref[fig2]{(a)}, the vertical blue dashed line indicates the critical Rabi coupling strength $\Omega_c^{\text{LDA}} \approx 0.126E_0$ predicted by LDA for $G = 1.3$.

In the more general case with non-zero spin-twisting angles ($\theta \ne 0^\circ$), the system develops a local spin polarization that is nonzero primarily near the edges. This edge-localized polarization cannot be captured by the integrated spin polarization alone. To characterize this effect, we employ the spin-scissors moment, which directly reflects the edge-dominated local spin polarization induced by the spin-twisting. After aligning the system and rescaling to make the spin-scissors mode dimensionless, it can be expressed as
\begin{equation}
    \mathcal{S}_\text{sc}=\frac{\langle xy \sigma_z \rangle}{\theta (R_{x0}^2-R_{y0}^2)},
\end{equation}
where $R_{x0}$ and $R_{y0}$ are the Thomas-Fermi radii of the system with the same parameters but without spin twisting. The interaction ratio $G \ne 0$ makes the two components repulsive to each other, which results in a tilt angle $\theta_\text{BEC}$ between two components large than the imposed spin-twisting angle $\theta$. When $\theta_\text{BEC}$ is small, we can get a simplified relation between $\mathcal{S}_\text{sc}$ and $\theta_\text{BEC}$,
\begin{equation}
    \label{thetaget}
    \begin{aligned}
            \theta_\text{BEC} &\approx \frac{\langle xy \sigma_z \rangle}{\langle x^2 \rangle_0 - \langle y^2 \rangle_0}\\
                    &\approx 6\theta \mathcal{S}_\text{sc},
    \end{aligned}
\end{equation}
where $\langle x^2 \rangle_0 = R_{x0}^2/6$ and $\langle y^2 \rangle_0 = R_{y0}^2/6$ correspond to the quadrupole modes $x^2$ and $y^2$ without the spin-twisting. For the time-dependent case, Eq.~\eqref{thetaget} also offers a straightforward description of spin-scissors-like dynamics, as will be discussed in detail in Sec.~\ref{sec4}.

To quantify the system's response to a spin-twisting, we introduce the spin-scissors susceptibility $\chi_s$, which describes the linear response of the spin-scissors mode. Specifically, $\chi_s$ is defined as
\begin{equation}
    \label{HD7}
    \chi_s = \lim_{\theta \to 0} \mathcal{S}_\text{sc}.
\end{equation}

In the phase-mixed regime, $\chi_s$ admits an analytical expression within the mean-field approximation. This expression can be derived under the LDA by minimizing the following energy functional~\cite{sartori2015spin}:
\begin{equation}
    \label{HD8}
    \mathcal{E} = \iint \left[\chi_{\text{sc}}^{-1} (n_1 - n_2)^2 + \eta xy (n_1 - n_2)\right] \, d\mathbf{r},
\end{equation}
where $\eta$ quantifies the strength of the spin-twisting effect and $\chi_{\text{sc}}$ is an effective local spin susceptibility, given by
\begin{equation}
    \label{HD9}
    \chi_\text{sc}(x,y) = \frac{2}{g - g_{12} + \Omega/n}.
\end{equation}
Minimizing the functional $\mathcal{E}$ with respect to the spin density imbalance $n_1-n_2$ yields the relation:
\begin{equation}
    \label{HD10}
    n_1 - n_2 = \eta xy\chi_{\text{sc}}(x,y).
\end{equation}
Substituting Eq.~\eqref{HD10} into Eq.~\eqref{HD7} leads to the analytical expression for the spin-scissors susceptibility:
\begin{align}
    \label{HD11}
    \chi_s &= \frac{m(\omega_y^2 - \omega_x^2)}{NR_x^2} \iint x^2 y^2 \chi_{\text{sc}}(x,y) \, d\mathbf{r} \nonumber \\
           &= \frac{1}{12}\frac{g + g_{12}}{g - g_{12}} \left[2 + f(\alpha)\right],
\end{align}
where $\alpha = \Omega / [n(x,y=0)(g - g_{12})]$, and the dimensionless function $f(\alpha)$ is defined as
\begin{equation}
f(\alpha) = 3\alpha(3 + 2\alpha) - 6\alpha(1 + \alpha)^2 \ln\left(1 + \frac{1}{\alpha} \right).
\end{equation}

\begin{figure}[t]
\centerline{
\includegraphics[width=0.45\textwidth]{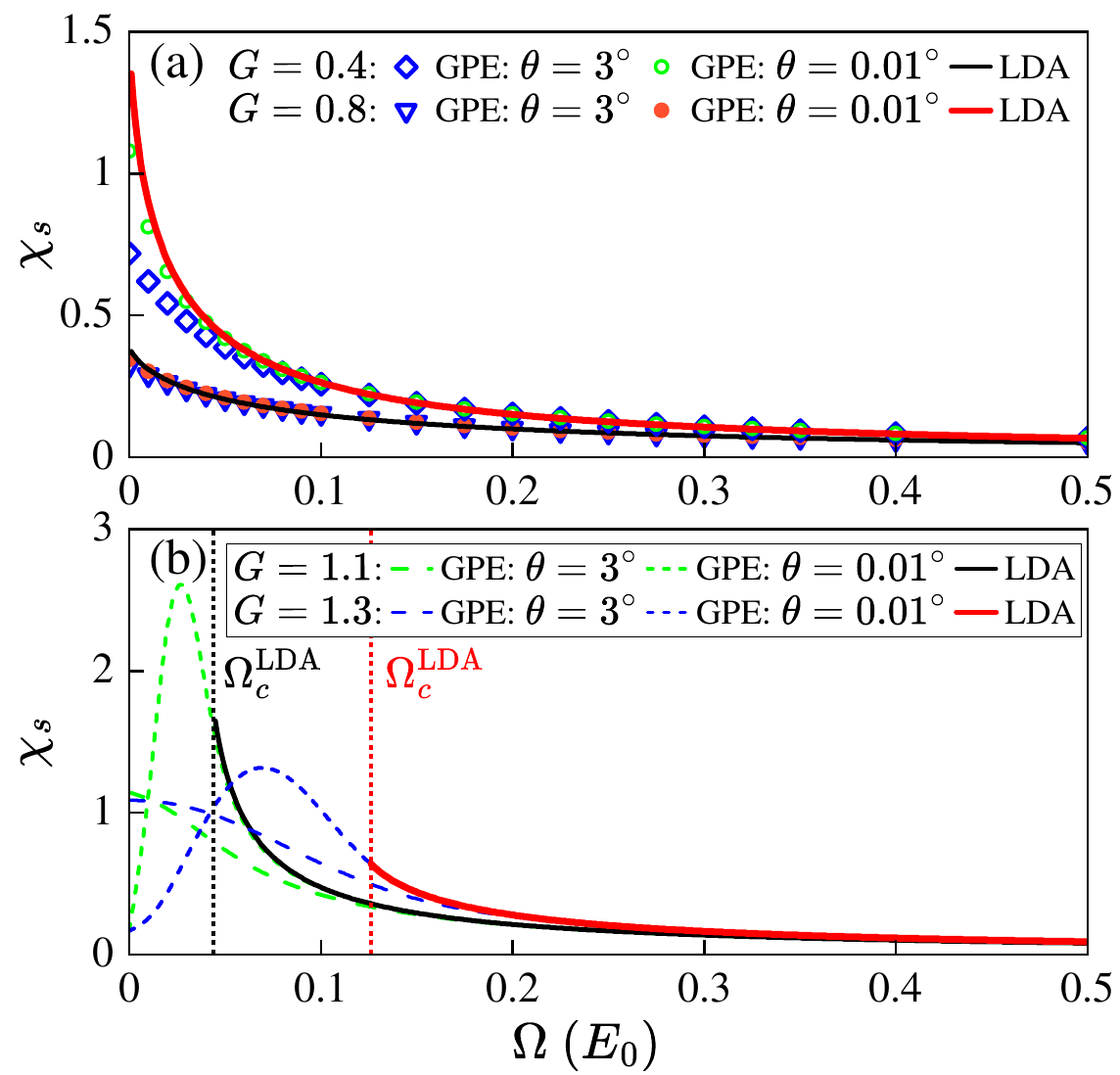}}
\caption{The dimensionless spin-scissors susceptibility $\chi_s$ as a function of the Rabi coupling strength $\Omega$. The GPE simulation results (colored symbols and dashed curves) for spin-twisting angles $\theta = 0.01^\circ$ and $\theta = 3^\circ$ are highlighted, while the LDA predictions (solid lines) which is valid when $\theta \to 0$ provide excellent agreement with the simulations. Panel (a) shows results for $G < 1$, and panel (b) for $G > 1$, with vertical dashed lines in (b) marking the critical Rabi coupling strengths $\Omega_c^\text{LDA}$.
}
\label{fig4}
\end{figure}

Figure~\ref{fig4} illustrates the behavior of the spin-scissors susceptibility $\chi_s$ as a function of the Rabi coupling strength $\Omega$ in two distinct interaction regimes, $G<1$ [Fig.~\ref{fig4}\hyperref[fig4]{(a)}] and $G>1$ [Fig.~\ref{fig4}\hyperref[fig4]{(b)}]. In both phase-mixed regimes with $(0\le G< 1,\,\Omega \ge 0E_0)$ [Fig.~\ref{fig4}\hyperref[fig4]{(a)}] or $(G>1,\,\Omega > \Omega_c^\text{LDA})$ [Fig.~\ref{fig4}\hyperref[fig4]{(b)}], increasing $\Omega$ enhances inter-component mixing, thereby suppressing the edge-dominated phase separation induced by the spin-twisting. As a result, the susceptibility $\chi_s$ decreases monotonically with increasing $\Omega$. In contrast, in the polarized regime for $(G>1,\,\Omega < \Omega_c^\text{LDA})$ shown in Fig.~\ref{fig4}\hyperref[fig4]{(b)}, the prediction of LDA breaks down; nevertheless, the GPE simulations still provide an accurate description of the system. In the nearly linear regime ($\theta = 0.01^\circ$), the susceptibility $\chi_s$ exhibits a pronounced peak at a finite value of $\Omega$ and tends to vanish as $\Omega$ approaches zero. This non-monotonic behavior reflects the suppression of the spin-scissors response due to the emergence of strong integrated spin polarization at weak Rabi coupling. At a moderate spin-twisting angle ($\theta = 3^\circ$), different interaction ratios $G$ yield values of susceptibility $\chi_s$ that are close to each other and remain nonzero as $\Omega$ approaches zero. A comparison of the results for $\theta = 0.01^\circ$ and $\theta = 3^\circ$ indicates that the spin-twisting-induced spin-scissors response at $\theta = 3^\circ$ is significant. Moreover, the comparison of different interaction ratios $G$ illustrates how the overall behavior of $\chi_s$ is sensitive to the relative strength of inter-component interactions. In the phase-mixed states, the results obtained from GPE simulations show excellent agreement with the predictions of the LDA.

\begin{figure}[t]
\centering
\includegraphics[width=0.45\textwidth]{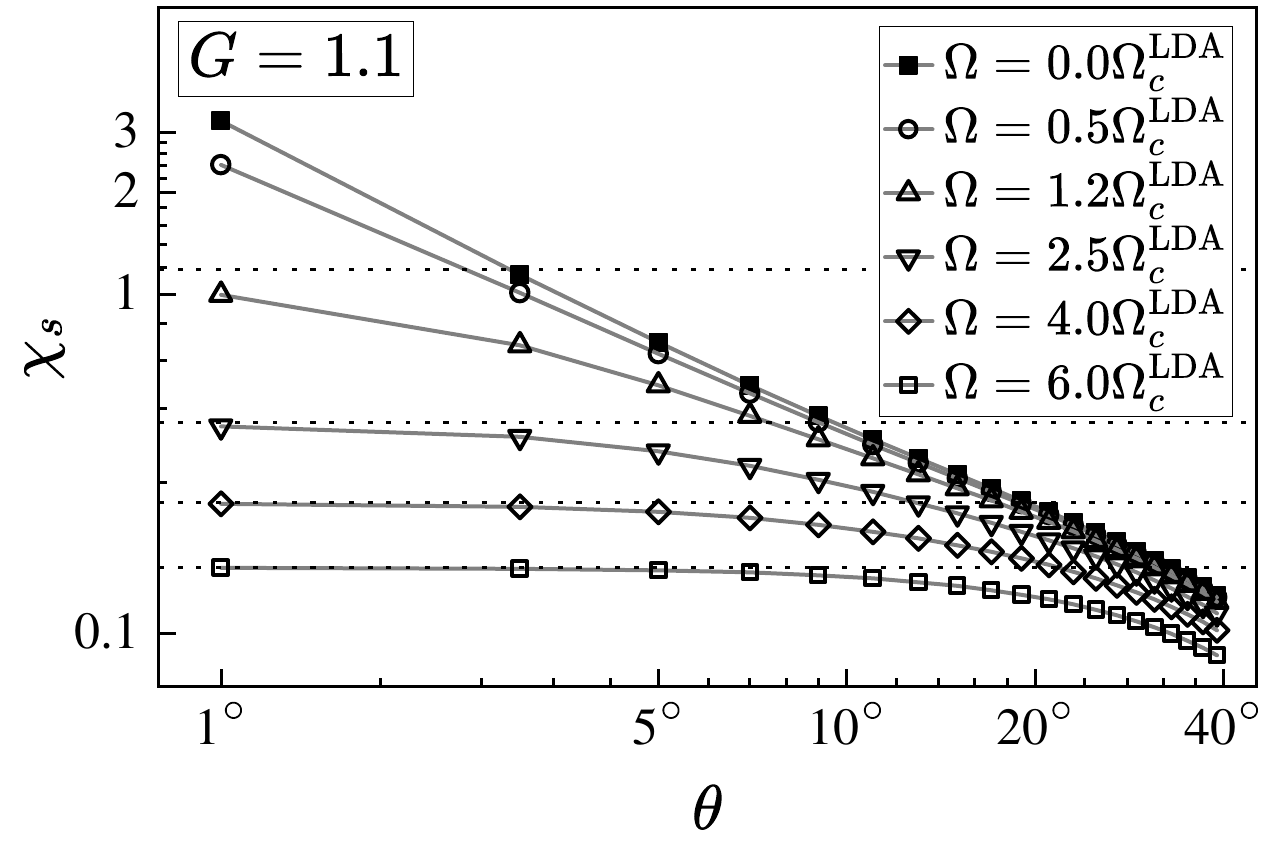}
\caption{The spin-scissors susceptibility $\chi_s$ as a function of the spin-twisting angles $\theta$, for different values of $\Omega / \Omega_c^\text{LDA}$ at fixed interaction ratio $G = 1.1$. The dashed lines correspond to the LDA results for four Rabi coupling strengths: $\Omega = 1.2, 2.5, 4.0,$ and $6.0\Omega_c^\text{LDA}$. The points represent the numerical results from GPE simulations. The gray solid line is included as a visual guide.
}
\label{fig5}
\end{figure}

To further elucidate the validity of the LDA description, we examine the dependence of the susceptibility $\chi_s$ on the spin-twisting angle $\theta$ for different values of $\Omega/\Omega_c^\text{LDA}$ in the regime with $G = 1.1$. As shown in Fig.~\ref{fig5}, the GPE results (pointed lines) and LDA predictions (horizontal dotted lines) from Eq.~\eqref{HD11} are compared over a wide range of $\theta$ from $1^\circ$ to $40^\circ$. When the Rabi coupling is sufficiently strong (e.g., $\Omega \gtrsim 6.0\Omega_c^\text{LDA}$), the system remains in a phase-mixed state, and $\chi_s$ exhibits only weak dependence on $\theta$ over a broader range of spin-twisting angles. In this region, the LDA yields quantitatively accurate predictions within the domain where the linear response approximation remains valid. However, as $\Omega$ decreases toward the critical value $\Omega_c^\text{LDA}$, deviations emerge due to the growing influence of nonlinear spin-density deformation beyond the linear response regime. In particular, the suppression of $\chi_s$ at large $\theta$ reflects the saturation of spin-twisting separation, which cannot be captured by the LDA based on small perturbations. These observations are fully consistent with the expected physical picture and validate the theoretical framework employed in this work.

\section{Spin-scissors Mode}
\label{sec4}

The spin-scissors mode was originally proposed as a theoretical refinement of the conventional nuclear scissors mode to incorporate spin degrees of freedom~\cite{PhysRevC.88.014306,PhysRevC.91.064312}. It manifests as a collective rotational oscillation described by spin motion. However, in nuclear experiments, clear signatures of this mode are often obscured due to low-energy background interference and overlapping spectral peaks~\cite{PhysRevC.97.044316,balbutsev2018nuclear}. Leveraging the controllability of ultra-cold atomics, we demonstrate that this elusive mode can be robustly excited via a sudden quench of the twisting angle between two spin components, thereby enabling the quantum simulation of spin-scissors dynamics.

In our two-component Bose-Einstein condensate system, the excitation protocol resembles that of the conventional scissors mode in single-component condensates. However, a key difference arises from the spin-dependent trapping potentials experienced by the two components, which lead to richer spin-related non-equilibrium dynamics. Specifically, the system is initially prepared with a small spin-twisting angle ($\theta = 3^\circ$) as shown in Fig.~\ref{fig1}, corresponding to a relative tilt of their density profiles. Abruptly turning off the twist excites the spin-scissors mode, which can be characterized by the time-dependent observable spin-scissors moment $\mathcal{S}_\text{sc}(t)$. The dynamics of the spin-scissors mode thus reflect the distinct properties of the underlying ground states.

\subsection{Regime $\bf 0\le G\le1$}

\begin{figure}[t]
\centering
\includegraphics[width=0.5\textwidth]{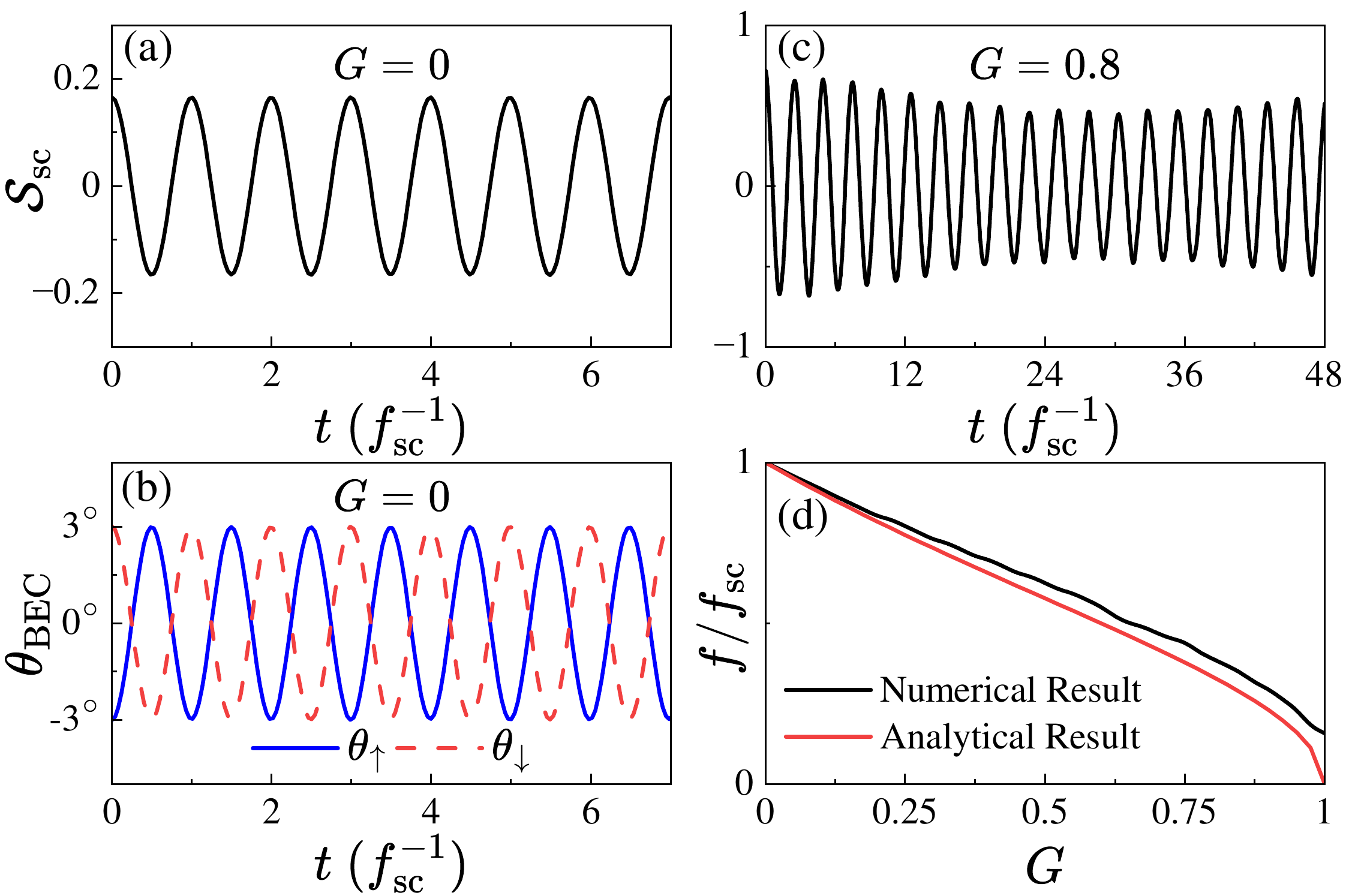}
\caption{(a–d) Spin-scissors mode dynamics without Rabi coupling ($\Omega = 0$). (a, c) Time evolution of the dimensionless spin-scissors mode $\mathcal{S}_\text{sc}$, shown for interaction ratios $G = 0$ (a) and $G = 0.8$ (c). The unit of time is set as the inverse of  scissors mode frequency $f_{\text{sc}} = \sqrt{f_x^2 + f_y^2}$. (b) Time evolution of the angular displacements $\theta_\uparrow(t)$ (blue solid line) and $\theta_\downarrow(t)$ (red dashed line) of the long axes of the two components relative to the $+y$ axis for $G=0$, calculated using Eq.~\eqref{thetaget}. (d) Spin-scissors mode frequency $f$ as a function of the interaction ratio $G$, in the absence of Rabi coupling. The black curve shows frequencies obtained from GPE simulations, while the red curve represents the analytical prediction $f/f_{\text{sc}} = \sqrt{(1 - G)/(1 + G)}$. All results correspond to an initial spin-twisting angle $\theta = 3^\circ$.
}
\label{fig6}
\end{figure}

As discussed in Sec.~\ref{sec3}, the system with spin-twisting remains in a phase-mixed state when $0 \le G \le 1$, characterized by zero integrated spin polarization but nonzero local spin polarization. During spin-scissors dynamics, each component maintains an elliptical TF profile, while the tilt angle $\theta_\text{BEC}$ of its major axis relative to the $+y$-direction evolves continuously in time. The dynamics of the spin-scissors mode are governed by both the inter-component interaction $G$ and the Rabi coupling $\Omega$. To isolate the effect of $G$, we first consider the case without Rabi coupling, $\Omega=0E_0$. Figures~\ref{fig6}\hyperref[fig6]{(a,c)} show the time evolution of the spin-scissors mode for two representative regimes: $G=0$ and $G=0.8$. For $G=0$, the two components oscillate independently; after turning off the initial twist, each component undergoes scissors oscillations at the intrinsic frequency $f_{\text{sc}} = \sqrt{f_x^2 + f_y^2}$ but in opposite phase, as further illustrated by the tilt-angle evolution in Fig.~\ref{fig6}\hyperref[fig6]{(b)}. In this case, the oscillation amplitude of $\theta_\text{BEC}$ equals the initial spin-twisting angle, $\theta=3^\circ$. When $0 < G \le 1$, the repulsive inter-component interaction couples the motions of two components. As a consequence, the tilt-angle amplitude of $\theta_\text{BEC}$ can exceed the initial twisting angle, $\theta=3^\circ$. This collision-mediated weak coupling further manifests in two key effects: a reduction (softening) of the spin-scissors frequency and a modulation of the oscillation amplitude. For example, in Fig.~\ref{fig6}\hyperref[fig6]{(c)}, the oscillation period is longer than in the uncoupled case of Fig.~\ref{fig6}\hyperref[fig6]{(a)}. Figure~\ref{fig6}\hyperref[fig6]{(c)} reveals a beating pattern in the dynamics for finite interaction ratios in the range $0 < G \le 1$, though without amplitude nodes. To characterize the frequency softening, we extract the fundamental frequency $f$ of the spin-scissors mode for varying $G$. We find that the spin-scissors oscillation frequency agrees very well with the following equation,
\begin{equation}
\label{fiteq}
f = f_{\text{sc}} \sqrt{\frac{1 - G}{1 + G}},
\end{equation}
which is reminiscent of the spin-dipole oscillation derived in Ref.~\cite{sartori2015spin}. Figure~\ref{fig6}\hyperref[fig6]{(d)} shows that, in the phase-mixed regime without Rabi coupling, the normalized frequency ($f/f_{\text{sc}}$) from time-dependent GPE simulations agrees with Eq.~\eqref{fiteq}, both exhibiting a consistent monotonic decrease with increasing $G$.

\begin{figure*}[t!]
\centerline{
\includegraphics[width=1\textwidth]{ 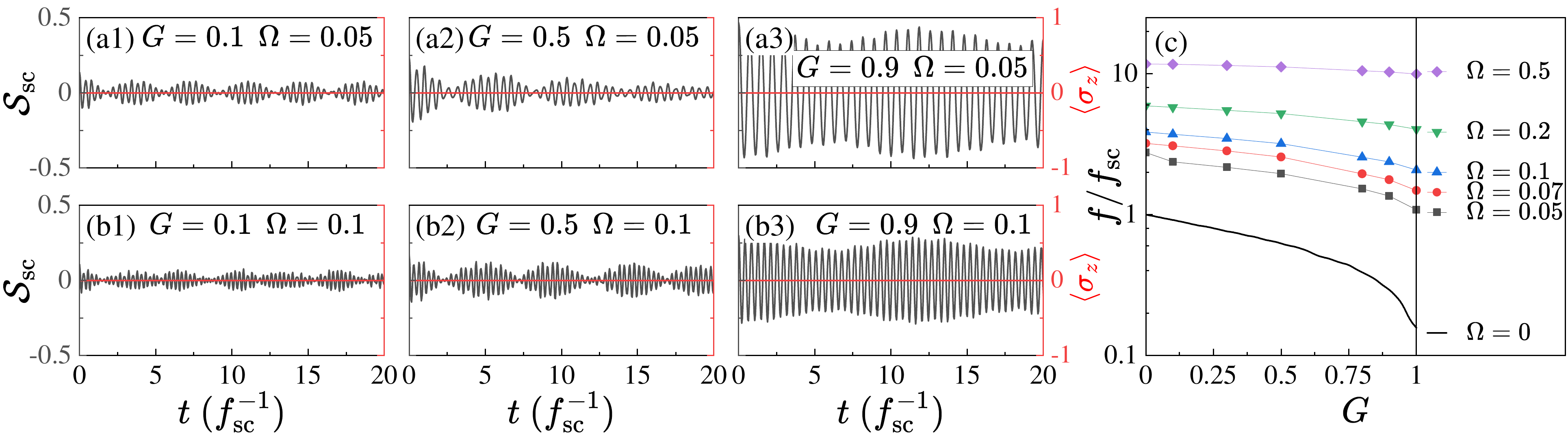}}
\caption{(a–c) Spin-scissors dynamics in the phase-mixed region at $0 \le G \le 1$ with nonzero Rabi coupling ($\Omega \ne 0E_0$). (a, b) Time evolution of the spin-scissors mode $\mathcal{S}_\text{sc}$ for Rabi coupling strengths $\Omega = 0.05E_0$ (a) and $\Omega = 0.1E_0$ (b). Each set shows results for three representative interaction ratios: $G = 0.1$ (a1, b1), $G = 0.5$ (a2, b2), and $G = 0.9$ (a3, b3). The red solid lines indicate the corresponding time evolution of the integrated spin polarization $\langle \sigma_z \rangle = 0$. The initial spin-twisting angle is fixed at $\theta = 3^\circ$. (c) Frequency $f$ of the spin-scissors dynamics as a function of the interaction ratio $G$ for various Rabi coupling strengths $\Omega$. The black solid curve reproduces the reference data for $\Omega = 0E_0$, as shown previously in Fig.~\ref{fig6}\hyperref[fig6]{(d)}.
}
\label{fig7}
\end{figure*}

The introduction of Rabi coupling ($\Omega \neq 0E_0$) induces strong coupling through spin exchange, in contrast to the weaker effect of the inter-component interaction $G$. The system remains in the phase-mixed regime with zero integrated spin polarization, $\langle\sigma_z\rangle=0$, throughout the dynamics. For small Rabi coupling ($\Omega=0.05E_0$) and weak inter-component repulsion ($G=0.1$), the spin-scissors mode exhibits a pronounced beating pattern together with a lager frequency $f$ than the fundamental frequency without Rabi coupling [Figs.~\ref{fig7}\hyperref[fig7]{(a1,c)}]. Increasing $G$ to $G = 0.5$ damps the beating and leads to a decrease of the frequency [Figs.~\ref{fig7}\hyperref[fig7]{(a2,c)}]. As $G$ approaches unity, the larger tilt angle $\theta_\text{BEC}$ suppresses node formation, so the beating disappears and the dynamics resemble those in the coupled case without Rabi coupling [cf. Fig.~\ref{fig6}\hyperref[fig6]{(c)}; Fig.~\ref{fig7}\hyperref[fig7]{(a3)}]. Under stronger Rabi coupling ($\Omega=0.1E_0$) [Figs.~\ref{fig7}\hyperref[fig7]{(b1-b3)}], $f$ exhibits an upward shift, while rapid spin-scissors oscillations distort the beating envelope. Figure~\ref{fig7}\hyperref[fig7]{(c)} summarizes the dependence of $f$ on $G$ for various $\Omega$: in all cases, $f$ decreases with increasing $G$, but the softening rate is progressively reduced as $\Omega$ increases. 

The pronounced and sustained beating pattern with well-defined nodes, shown in Fig.~\ref{fig7}\hyperref[fig7]{(a1)}, results from resonance between the fundamental mode frequency $f_1$ [at $G=0.1,\ \Omega=0E_0$; black curve in Fig.~\ref{fig7}\hyperref[fig7]{(c)}] and the Rabi coupling frequency $f_2 = \Omega/(2\pi\hbar)$ at $\Omega=0.05E_0$, such that $f_1 \approx f_2$. In contrast, under other parameter configurations, the dynamics exhibit damped, irregular, or node-deficient oscillations due to off-resonance ($f_1 \ne f_2$). For instance, at $\Omega=0.5E_0$, where $f_2 \gg f_1$, the dynamics are dominated by Rabi oscillation.

In the regime $G<1$, the spin-scissors dynamics mainly reflect the generic features of the phase-mixed state, characterized by stable, periodic modulations. This behavior contrasts with the phase-mixed regime at $G>1$ and $\Omega>\Omega_c^\theta$, where the dynamics exhibit spontaneous spin polarization. The following subsection will be devoted to a detailed discussion of this phenomenon.

\subsection{Regime $\bf G > 1$}

\begin{figure*}[t!]
\centerline{
\includegraphics[width=1\textwidth]{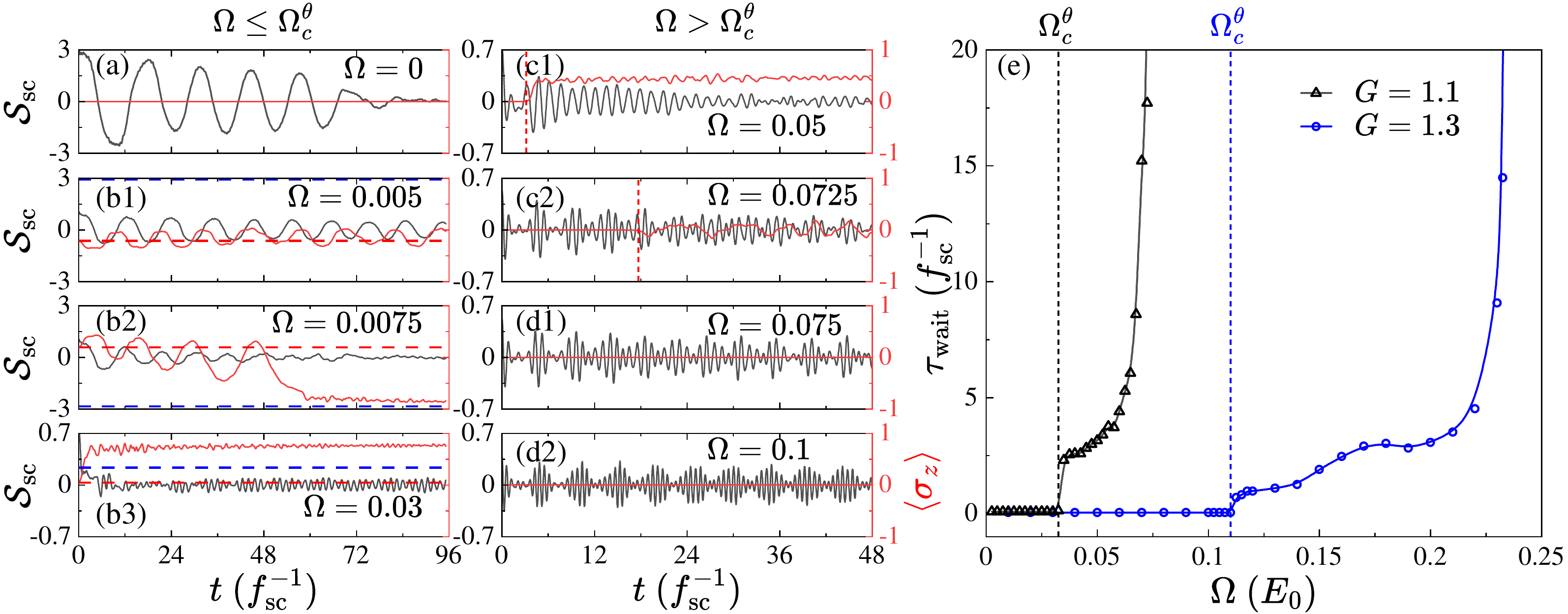}}
\caption{(a–d) Representative time evolutions from GPE simulations for interaction ratio $G = 1.1$. The black solid curves show the spin-scissors mode $\mathcal{S}_\text{sc}$, and the red solid curves show the corresponding integrated spin polarization $\langle \sigma_z \rangle$. In (b), the horizontal red and blue dashed lines indicate the integrated spin polarization of the initial state and the post-quench ground state, respectively. In (c), vertical red dashed lines mark the waiting time $\tau_{\text{wait}}$, extracted from each evolution as the first noticeable deviation of $\langle \sigma_z \rangle$ from zero; these values correspond to the data points in (e). (e) Waiting time $\tau_{\text{wait}}$ as a function of the Rabi coupling strength $\Omega$, defined as the time interval between the quench and the onset of spontaneous polarization. Symbols represent GPE results for different interaction ratios: black triangles for $G = 1.1$, and blue circles for $G = 1.3$. Vertical dashed lines in corresponding colors indicate the critical Rabi coupling strength $\Omega_c^\theta$ for each value of $G$ under the condition of the spin-twisting angle $\theta=3^\circ$.
}
\label{fig8}
\end{figure*}

From the analysis of ground state in Sec.~\ref{sec3}, the Rabi coupling $\Omega$ promotes global uniformity across the system, whereas the inter-component interaction with $G>1$ favors the center-driven phase separation. Moreover, applying spin-twisting effectively promotes the edge-dominated phase separation and local spin polarization with suppressing the integrated spin polarization. In the regime $G>1$, the system supports three distinct phases, phase-separated, polarized and phase-mixed phases. In contrast to the 1D spin-dipole dynamics reported in Ref.~\cite{sartori2015spin}, which also address the $G>1$ regime, the 2D spin-scissors dynamics studied in this section shows qualitatively different behavior under comparable conditions. As $ \Omega $ varies from $ \Omega=0E_0 $ to the point $ \Omega\to\infty $, the system exhibits four distinct dynamical behaviors. Figures~\ref{fig8}\hyperref[fig8]{(a-d)} show these four cases for $G = 1.1$—a value chosen due to its relatively small $\Omega_c^\text{LDA} \approx 0.044 E_0$, which leads to slower dynamical frequencies as seen in Fig.~\ref{fig7}\hyperref[fig7]{(c)} and enables clearer visualization of the behavior.

(I) Figure~\ref{fig8}\hyperref[fig8]{(a)}: With no Rabi coupling ($\Omega=0E_0$) and $G>1$, the system starts from a phase-separated state. Integrated spin polarization remains zero throughout the dynamics. After a few long-period oscillations (on the order of $f_\text{sc}$), the two components stabilize and the spin-scissors mode $\mathcal{S}_\text{sc}$ vanishes. (II) Figures~\ref{fig8}\hyperref[fig8]{(b1-b3)}: For $0<\Omega\le\Omega_c^\theta$, the system lies in the polarized regime. As $\Omega$ increases, three behaviors emerge: (i) at small $\Omega$, oscillations occur around the initial polarization [see~\hyperref[fig8]{(b1)}]; (ii) at intermediate $\Omega$, oscillations are gradually suppressed and the system stabilizes near the post-quench ground-state polarization [see~\hyperref[fig8]{(b2)}]; and (iii) approaching $\Omega_c^\theta$, where the initial polarization vanishes, oscillations are abruptly quenched, but the steady-state polarization deviates from the post-quench value [see~\hyperref[fig8]{(b3)}]. (III) Figures~\ref{fig8}\hyperref[fig8]{(c1,c2)}: For $\Omega>\Omega_c^\theta$, starting from a phase-mixed state with zero polarization, the system maintains $\langle\sigma_z\rangle=0$ for a finite waiting time $\tau_\text{wait}$ before developing spontaneous polarization. With increasing $\Omega$, $\tau_\text{wait}$ grows significantly while the final polarization diminishes. (IV) Figures~\ref{fig8}\hyperref[fig8]{(d1,d2)}: For $\Omega\gg\Omega_c^\theta$, $\tau_\text{wait}$ diverges and the system exhibits a stable beating pattern of the spin-scissors mode, characteristic of the phase-mixed regime discussed previously.

Figure~\ref{fig8}\hyperref[fig8]{(e)} shows the waiting time $\tau_\text{wait}$ versus Rabi coupling $\Omega$ for interaction strengths $G=1.1$ and $1.3$. After the critical Rabi coupling $\Omega_c^\theta$, $\tau_\text{wait}$ exhibits an abrupt rise from zero, marking the phase transition from polarized to phase-mixed states. This sharp increase originates from the mismatch of integrated spin polarization between the initial unpolarized phase-mixed state and the polarized post-quench ground state [cf.Fig.~\ref{fig2}\hyperref[fig2]{(a)}]. Maintaining the unpolarized configuration of the initial state requires a finite time, leading to the onset of long waiting times. Moreover, the abruptness of the rise diminishes with increasing $G$, since the polarization mismatch is reduced, so the rise for $G=1.3$ is weaker than that for $G=1.1$. Comparing the two cases further reveals that significant divergence of $\tau_\text{wait}$ occurs in the region $2\Omega_c^\theta < \Omega <3\Omega_c^\theta$, where the range of spontaneous polarization is narrow for $G=1.1$ but broadens considerably for $G=1.3$. This demonstrates that the interaction strength $G$ effectively controls the dynamical miscibility window of the system.

It is worth noting that a similar waiting time $\tau_\text{wait}$ has been used to characterize spin-dipole dynamics in 1D systems~\cite{sartori2015spin}, where it quantifies how long the system remains unpolarized before spontaneously polarization. However, the qualitative behavior differs: in 1D spin-dipole dynamics, $\tau_\text{wait}$ decreases monotonically with increasing $\Omega$, whereas in 2D spin-scissors dynamics it increases and even diverges at large $\Omega$. Despite this contrast in the evolution of $\tau_\text{wait}$, the degree of spontaneous polarization itself decreases monotonically with increasing $\Omega$ in both cases. These properties highlight that different geometries can lead to distinct dynamical mechanisms.

\section{Conclusions}
\label{sec5}

Within the LDA analysis and GPE simulations, we systematically investigate the ground-state properties and spin-scissors dynamics of binary BECs in a twisted 2D harmonic trap. The ground state exhibits three distinct phases—phase-separated, polarized, and phase-mixed—determined by the Rabi coupling $\Omega$, the interaction ratio $G$, and the spin-twisting. The spin-twisting introduces a position-dependent detuning under the linearized approximation, favoring edge-dominated phase separation and thereby generating local spin polarization at the condensate edges. Moreover, the ground state exhibits a finite spin-scissors susceptibility $\chi_s$, with numerical GPE simulations showing excellent agreement with LDA predictions in the phase-mixed regime.

For $0 \leq G \leq 1$, the system remains phase-mixed with zero integrated spin polarization. In this regime, inter-component repulsion induces frequency softening and amplitude modulation, while Rabi coupling can drive the persistent beating pattern with enhanced frequency. These dynamics are stable and periodic, in sharp contrast to the spontaneous polarization that arises for $G>1$. For $G>1$, the system exhibits four distinct dynamical regions as $\Omega$ increases: (i) beat damping in the phase-separated state at $\Omega = 0E_0$; (ii) polarized relaxation in the polarized state for $0 < \Omega \leq \Omega_c^\theta$; (iii) phase-mixed states with a finite waiting time when $\Omega > \Omega_c^\theta$; and (iv) stable beating at large $\Omega$. A key feature is the emergence of spontaneous polarization after a waiting time $\tau_\text{wait}$, whose growth and divergence reflect the polarization mismatch between initial state and post-quench ground state. Compared with 1D spin-dipole dynamics, where $\tau_\text{wait}$ decreases monotonically with $\Omega$, the 2D spin-scissors dynamics reveal a qualitatively different mechanism controlled by geometry and interaction ratio $G$. These dynamical responses evolve smoothly with increasing $\Omega$, collectively outlining a unified picture that connects ground-state properties to non-equilibrium dynamical behaviors.

\begin{acknowledgments}
This work was supported by the National Natural Science Foundation of China (NSFC) under Grant Nos. 12374247 and 11974235, and by the Shanghai Municipal Science and Technology Major Project (Grant No. 2019SHZDZX01-ZX04). X.Z. acknowledges support from the Zhejiang Provincial Natural Science Foundation of China (Grant No. LD24F020009). C.Q. is supported by ACC-New Jersey under Contract No. W15QKN-18-D-0040.
\end{acknowledgments}


\input{main_for_arxiv.bbl}

\end{document}

%% file: main_for_arxiv.bbl
%